# Meta-analysis from different tracers of the small Local Arm around the Sun - extent, shape, pitch, origin

Jacques P. Vallée

National Research Council of Canada, Herzberg Astronomy & Astrophysics, 5071 West Saanich Road, Victoria, B.C., Canada V9E 2E7

**Abstract.**

The Sun in not located in a major spiral arm, and sits in a small 'Local Arm' (variously called arm, armlet, blob, branch, bridge, feather, finger, segment, spur, sub-arm, swath, etc). The diversity of names for the 'Local Arm' near the Sun indicates an uncertainty about its shape or pitch or its extent from the Sun in each galactic quadrant, as well as an uncertainty about its origin.

Here we extract data about the small 'Local Arm' near the Sun, from the recent observational literature, over many arm tracers, and we use statistics in order to find the local arm's mean extent from the Sun, its possible shape and pitch angle from the direction of galactic longitude 90°. Employing all tracers, the Local Arm is about 4 kpc long by 2 kpc large. The Sun is within 1 kpc of the center of the local arm. Proposed 'bridges' and 'fingers' are assessed. These bridges to nearby spiral arms and fingers across spiral arms may not reach the nearest spiral arms, owing to kinematic and photometric distance effects.

We then compare these statistical results with some predictions from recent models proposed to explain the local arm (perturbations, resonances, density wave, halo supercloud, debris trail from a dwarf galaxy).

The least controversial models involve importing materials from elsewhere (halo supercloud, debris trail) as a first step, and to be later deformed in a second step (by the Galaxy's differential rotation into become roughly parallel to spiral arms) and then subjected to ongoing forces (global density waves, local perturbations).

## 1. Introduction

A recent review of the locations of the four well-known spiral arms in our Milky Way galaxy was given in Vallée (2017c).

Distances to local stars obtained through optical observations are variously affected by *dust*. Incomplete or no removal of dust effect make stars appear more distant, while they could be closer to the Sun in reality.

Also, using an incomplete velocity-distance relation, observations of the velocity *dispersion* could transform a stellar cluster into a radially elongated feature, causing the appearance of a 'finger' going away from the Sun.

Here, a statistical analysis is made to give a better precision to its shape, direction, and size, and to compare with some theoretical models as previously proposed. Section 2 deals with individual radio and optical tracers, such as O-B stars, Cepheid stars, masers, etc (Table 1; Table 2). Section 3 deals with mapping of the tracers into the Galactic plane (Figure 1, Figure 2, Table 3). Section 4 deals with azimuthal bridges between arms, as proposed in the literature (Figure 3). Section 5 deals with radial fingers crossing arms, as proposed in the literature (Figure 3). Section 6 deals with possible origins for the small local arm. Section 7 provides a concluding discussion.

## 2. Extent, shape, pitch according to individual tracers

Various tracers have been used to delineate the local arm, notably for its shape, pitch, and extent. Many tracers only have a rough measured distance from the Sun, not better than 10% or more.

**Table 1** shows some representative results, involving Cepheids groups, HI gas, HII regions, molecular clouds, O-B stars, open clusters, pulsars, and trigonometric masers. Half of the data in Table 1 are radio tracers with a precise distance (about 3%), while the remainders are optical tracers.

Extent-wise, none of the tracers in Table 1 show a distance beyond 4 kpc.

Shape-wise, some of these tracers showed a random distribution (Cepheids Groups, open clusters, pulsars). Others are ambivalent, depending on the methodology (molecular clouds, O-B stars). Others appear to show an elongation (HII regions, HI gas, masers).

Pitch-wise, none of the tracers in Table 1 seem to agree much in pitch angle, as there is a widespread range for the pitch angle (from +3 for pulsars, to -30 for open star clusters). Most tracers show a random pattern in pitch angle.

The only tracer with a well defined pitch is the masers, each having a well measured trigonometric distance with a distance error better than 3%.

**Table 2** shows the results of all the observed trigonometric masers: galactic longitude, latitude, solar distance, radial velocity, and reference. All the data in Table 2 are radio tracers with a precise distance.

## 3. Extent, shape, pitch according to statistics

**Table 3** shows some statistical means, done on the data in Tables 1 and 2.

In Table 3, statistics are performed separately for the trigonometric masers (best distance estimates). Statistics are also done on all the other tracers (kinematical or photometric distance estimates) with a common weight.

**Figure 1** shows the size of the Local Arm, from the data in Table 1, draw as four long dashes joined together. The trigonometric masers are also shown, as 'x' signs. The Sun is shown as a star at (8,0, 0.0).

For the four long arms, the arm model shown in Fig.1 is that of Vallée (2017b – his Fig.5a; 2017d – his Fig.3). It employed the basic parameters of a Sun to Galactic Center of 8 kpc (Vallée 2017a), mean arm pitch of -13.1° (Vallée 2017b), and logarithmic shape (Vallée 2017c). The model fits very well the arm tangents (galactic longitudes) in the CO tracers for Sagittarius-Carina, Scutum and Norma (catalogued in Tables 3 to 10 in Vallée 2016b).

Using all tracers except the trigonometric masers, the bottom of the local arm (toward l=0°) has been cut at the location of the Sagittarius arm; there is some confusion in the literature as to where exactly the local arm encounters the Sagittarius arm (Fig. 7 in Chené et al 2013; Fig. 10 in Genovali et al 2014).

Extent-wise, excluding trigonometric masers, one can see in Table 3 that the Local Arm has a mean width not exceeding about 2.3 kpc from the Sun in Galactic quadrants (GQ) I and III, and not exceeding 1.9 kpc in the other quadrants. The extent of the 'Local Arm' in Figure 1 in the galactic plane (all tracers, excluding the trigonometric masers) is about 4 kpc (horizonthal) by 2 kpc (vertical), and its vertical component may have been corrupted by stars close to the Perseus arm (not really belonging to the Local Arm). Excluding trigonometric masers, the Sun's very little separation from the small local arm center is similar to that found in Fig. 2a of Liu et al (2017) for young F-type stars (about 0.2 kpc in the azimuthal direction), and in Fig. 7 in Du et al (2017) for the $H_2$ gas (about 0.6 kpc cos[145° - 90°], or 0.3 kpc).

Extent-wise, using only the trigonometric masers, one sees a local arm's extent near 5 kpc, and the Sun's location about 2 kpc from the local arm's center, a result similar to that found in Fig. 5 in Hu et al (2016). This extent is close to the extent of the 'swath' of about 5 kpc (horizonthal) by about 1 kpc (vertical) as proposed in Xu et al (2013, their Fig. 10 and Section 4).

Shape-wise, it resembles an ellipsoidal 'blob', with a major axis about 4 kpc oriented with a pitch near -10°, thus pointing toward a galactic longitude l near -80°.

Pitch-wise, one can see that the trigonometric masers have a mean pitch near -16° with an extent from the Sun of about 2 kpc in GQ III and 3 in GQ I.

Comparisons. Within 1.5 kpc of the Sun (circle in figure 1), recent observational data match well with our Figure 1, notably: (1) optical interstellar dust - Gaia data (Fig. 7 and Fig. 8 in Lallement et al 2018); (2) O-type stars near the Sun – Gaia data (Fig. 2 in Xu et al 2018); (3) molecular clouds (Fig. 11 in Miville-Deschênes et al 2017).

Looking towards the Galactic Center (l=0º) the line-of-sight meets the location of the Sagittarius mid-arm in the CO tracer at 0.9 kpc from the Sun (Fig.1 here; Fig.4 in Vallée 2017d). Given the offset between CO and masers, with a mean of about 0.32 kpc (Fig.1 in Vallée 2016b), then the maser lane at the inner edge of the Sagittarius arm is at 1.1 kpc from the Sun; this compares well with the maser observations (Fig.1 in Reid et al 2014), with proper account for the different distances of the Sun to the Galactic Center.

**Figure 2** shows the locations in radial velocity space of the trigonometric masers toward the anti-Galactic Center region (GQ II and III). All are within about 20 km/s from the Sun's radial velocity of 0 km/s.

The arm model shown in Figure 2 is that of Vallée (2017b – his Fig.5b, for GQ II and II), and of Vallée (2017d – his Fig.4, for GQ IV and I). An orbital circular speed of 230 km/s is employed (Vallée 2017a), as well as the start of each spiral arm near 2.2 kpc (Vallée 2017b). This recent velocity model improved on competing models (Table 3 in Vallée 2017b).

### 4. Bridges, as proposed in the literature

So-called 'bridges' (short features going very roughly parallel to a long arm) near the Sun have been proposed and published elsewhere (shown here in orange dashed lines).

In **Figure 3**, in GQ I, we show the 'bridge' to the Sagittarius arm from Fig. 2 in Xu et al (2016), and the model 'bridge' (no tracers) to the Perseus arm from Fig. 12(a) in Xu et al (2013). In GQ III, we show the 'stellar bridge' to the Perseus arm, near l=258º from Fig. 15 in Giorgi et al (2015). In GQ IV, we copy the 'fork' to the Carina arm as sketched in Fig. 15 of Carraro et al (2017) as a 'connection to spiral arms' (the open clusters are not necessarily located in the fork, but not far from it).

All 'bridges' are proposed models, drawn by starting from more nearby observational data, and assuming a line being projected out to far away. At some places in the linear model projections, there are many gaps without observed data. There seems to be a real lack of stars in between, so this would make the bridge unphysical (more like a 'constellation' as seen in the sky).

Two earlier proposed complex bridges near the Sun (not shown here), going from the Sagittarius arm at l=0º to the Perseus arm at l=90º, were modeled onto local stellar velocity data and local HI gas data (Fig.3 in Englmaier et al 2011). Both these models now contradict the well-determined observed tangents to the Sagittarius and Carina spiral arms, as measured in many tracers (Vallée 2016b).

Several new photometric distances to young open clusters appear to be smaller than the older ones by a factor of about 2 (Table 6 and Figure 15 in Carraro et al 2017), due to anomalous reddening and extinction. The new distances put most of the young open clusters inside the width of the Sagittarius arm, when plotting Fig. 4 in Vallée (2016a) over Fig. 15 of Carraro et al (2017).

### 5. Fingers, as proposed in the literature

So-called 'fingers' (long features radially elongated along the line of sight) have been published elsewhere (shown here in blue dashed lines in Figure 3).

There is one for the 4-kpc line of O stars in Fig. 4 in Bobylev and Bajkova 2015 near $l=190°$; one for the 10-kpc line of CO clouds in Fig. 15 in Vazquez et al 2010 near $l=260°$; one for the 9-kpc line of young open star clusters in Fig.1 in Carraro et al 2015 near $l=240°$; ditto for open clusters in Fig. 4 in Giorgi et al 2018 near $l=240°$; and those for the 9-kpc line of HI toward $l=200°$, the 9-kpc line towards $l=235°$ and the 7-kpc line towards $l=150°$ in Fig. 4a in Levine et al (2006).

All 'fingers' are proposed models, drawn from nearby data and projected far away. At some places in the projections, there are spatial gaps without observed data. These fingers do not appear to interact with the Perseus spiral arm in GQ II and III. There seems to be a real lack of data in between, so this would make the fingers unphysical (sometimes labelled 'fingers of God').

### 6. Origins for the small local arm

How do we explain the presence of a Local Arm, and its dynamics?

Various theoretical models have been proposed, some employing local matter (to be perturbed or re-organized), and some importing new matter (from the galactic halo, or a dwarf galaxy). It is possible that, over time, one model is better for the beginning of a local arm, and another model then takes over later on to shape it as it is now observed.

### 6.1 Perturbations - not a simple bridge between two nearest spiral arms

Each individual model proposed in the literature is simple and often localized in its own Galactic quadrant. But, taking all models together to cover all Galactic Quadrants, explaining so many bridges and fingers near the Sun becomes difficult and contradictory.

In Figure 1, in GQ IV, one sees that the 'fork-like' structure already covers part of the Carina spiral arm (the name of the Sagittarius arm in that quadrant).

Theoretically, some kind of local but strong perturbation could have occurred in a major arm, separating it into a 'fork'. In GQ IV, theoretically a small portion of the Sagittarius arm may have separated and then become elongated by the galactic differential rotation.

Observationally, pitch-wise, the low pitch angle of this local arm, near -10º (Table 2), is similar to the pitch angle of the major arms, and this pitch is too small to quickly reach either the Perseus arm or the Sagittarius arm; in other galaxies, some bridges have a higher pitch angle.

Extent-wise, the short length of this local arm, near 3 kpc (Table 1), is not long enough at this low pitch angle to reach the nearest spiral arms.

Shape-wise, if all these bridge models are real, then somehow around the Sun there is a complex 'staircases' linking the Sagittarius arm to the Perseus arm, in the form of an elongated x-type structure.

Using the observed thermal electron distribution, some have proposed to shift inward a 3-kpc segment of an existing long spiral arm, namely the Sagittarius arm between l=0º and l=45º, giving it a flat line (not a logarithmic shape), as in Fig. 9 in Cordes and Lazio (2003).

Given the observational gaps in observational data (Sections 4 and 5), and the short extent of the Local Arm, the theoretical case for a local perturbation able to separate (fork in GQ IV) or shift (displace in GQ I) a segment of the existing Sagittarius-Carina arm has yet to be proven.

**6.2 Mixed tracers - not a long density wave, and not near co-rotation**

*Perpendicular* to the local arm, one sees that the width of the trigonometric masers is about 1 kpc, much larger than expected from density wave theory. Also, there is no offset between different tracers (CO, dust, masers) as found in density-wave spiral arms; it is observed to be about 315 pc between dust, masers versus broad CO tracers– see Fig. 1 in Vallée (2014) or Tab. 1 in Vallée (2016b).

*Parallel* to the small local arm, its observed length is too short to be the product of the density-wave theory; it is not a galactic-wide arm, contradicting predictions of a long arm (e.g., Griv et al 2019).

Shape-wise, the local arm's distance to the Sagittarius arm is only about 1 kpc, much shorter than the 3 kpc distance needed to go from the Sagittarius arm to the Perseus arm in a regular spiral arm model (Vallée 2014 – his Fig. 2).

The Local Arm is at a radial distance of 8 kpc from the Galactic Center, while the corotation radius (gas and density wave going at the same speed) was shown to be located beyond the Perseus arm (nearer 12 kpc; Sahai et al 2015; Vallée 2018). For a corotation nearer 12 kpc, the Sun's location is near the 4:1 Inner Linblad resonance, thus perturbing the small Local Arm.

The observed length of the Local Arm is not long enough in GQ III to show the 'banana-like' shape and its 8-kpc extent as predicted if located near the co-rotation radius (Lépine et al 2017). The Hercules stream near the Sun has been analysed as a 'possible' signature for the 4:1 Outer Linblad Resonance from a slow and long bar near the Galactic Center (Hunt and Bovy 2018), and also as a 'possible' signature of the 8:1 Inner Lindblad resonance (Michtchenko et al 2018).

The physical complexity of the Local Arm (Fig. 1) may perhaps be linked to the velocity complexity observed in the Gaia DR2 data; thus the plots of the azimuthal velocity versus radial velocity of Gaia stars have revealed a dozen 'arches' (Table 2 and Fig. 3 in Ramos et al 2018). An exact one-on-one link is not possible as Gaia deals with optical stars while radio masers are not yet emitting in the optical.

Another small local arm was found between the Scutum arm and the Sagittarius arm, thus still inside corotation, near l=32º to 39º, and $v_{lsr}$= +60 to +90 km/s (Rigby et al 2016) – about 5 kpc from the Galactic Center, below the corotation radius.

### 6.3 A possible supercloud, from the galactic halo

The presence in the Local Arm of several stellar groups and stellar streams is well known (Hercules stream, Gould's Belt, Cygnus complex, etc). What created those streams, and where did they come from?

The mass of the Local Arm is not well known, but apparently greater than a typical interarm region between two inner arms. The Cygnus Complex (towards galactic longitude l=80º, at a distance of about 1.4 kpc) was estimated at near $10^7$ $M_{sun}$ (Tibaldo et al 2011). Assuming ten such complexes within 3 kpc of the Sun, the total mass of the Local Arm would be at least $10^8$ $M_{sun}$. Bigger complexes, such as the Monoceros Ring and the Helmi stream, both near the Sun, are each one estimated to be near $10^8$ $M_{sun}$.

Olano (2016, his Fig.5) proposed a model of a $10^7$ $M_{sun}$ supercloud passing through a part of the Perseus arm (100 Myrs ago); this passage was braked, creating a shock and a collapse of its inner part (to become Gould's Belt), as well as an expansion of its outer part (to be deformed by the Galaxy's differential rotation), and curving its orbital path to approach the Sagittarius-Carina spiral arm (now).

Thus the disintegration of the supercloud could form new subsystems (Local arm, Gould's belt), near the Sun (his Fig. 12).

Nogochi (2018) also proposed two-accretion events near the Sun from the Milky Way's halo (infalling gas streams, separated by 2 billion years).

### 6.4 A debris trail from a dwarf galaxy's successive orbital passages

Some dwarf galaxies have come close to the Milky Way's disk in the past. Should an earlier such small dwarf galaxy left debris near the Sun in the past, then some spur is expected to be present in the galactic disk. The incoming stuff may be easier seen in between the arms, rather than in the existing arms. At each of several passages, such a dwarf galaxy may shed mass in a trail. After each passage, the dwarf galaxy's mass will decrease, while the interarm mass would increase at specific locations.

For example, the current close Sagittarius Dwarf galaxy (SDG) is at a mere 16 kpc from our Milky Way's Galactic Centre, toward longitude l= 5º and latitude b= -15º, and radial velocity near +140 km/s, with a size of about 3 kpc – see Fig. 3 in Ibata et al (1994). A recent passage of SDG, with a mass near $10^8$ $M_{sun}$, crossed the Milky Way's disk between the Perseus arm and the Cygnus arm; its trailing debris stream perpendicular to its orbital plane has a width of about 2.0 kpc – see Fig. 1 in Law et al (2005). In the future, SDG may possibly add some material in the Milky Way's disk (a small fraction of its mass). That trailing material may be affected by the Milky Way's differential rotation (elongating the new spur).

The Gaia DR2 optical stellar velocity data were analysed by Antoja et al (2018) who found many velocity substructures (arches, shells, ridges) and effects from a recent passage of the SDG.

### 7. Concluding discussion

We investigated numerous published results on the Local Arm, notably maximum extent from the Sun, shape and pitch angle. The advent of Gaia DR2 has not changed this picture much, given that the optical stars in the Local Arm are already close to the Sun.

Observationally, our main results are:

1-the Local Arm is real but small, as seen in many different tracers close to the Sun, not a long global arm;

2-most tracers (in Table 1) indicate a random distribution, while the trigonometric masers (Table 2) show a non-random distributions;

3-many optical tracers are centered close to the Sun (Table 3, Figure 1), not exceeding a small distance of about 1.5 kpc;

4-the trigonometric masers indicate a small azimuthal Local Arm, of about 4 kpc long by 2 kpc large;

5-the radial velocity of the trigonometric masers is close to that of the Sun (Fig. 2);

6-some tracers look like short 'bridges' or long 'fingers' (Fig. 3), yet these bridges have *not* reached the long normal spiral arms, while these fingers do *not* seem to interact with normal long arms (Perseus, Sagittarius).

Theoretically, our main results are:

1-there is no plausible theoretical model of a simple *long* spiral arm to describe the small local arm;

2-we find serious objections to simple density waves and to simple model bridges and simple model fingers;

3-looking across the local arm, we find no offset between tracers as found in density-wave spiral arms – observed for the Sagittarius and Scutum arms to be about 315 pc between dust/masers versus cold broad CO tracers;

4-the large mass of the Local Arm, contained within 3 kpc, suggests a more recent origin than the older origin of the four long spiral arms;

5-this enhances the possibility that the 'Local Arm' is a complex phenomenon, with a beginning from an orbiting supercloud or a debris trail from an orbiting dwarf galaxy to *import* the material (or both).

6-the Local Arm can be though as a collection of several groups, each added at a different time, and shaped or re-shaped by several physical phenomena over time (Arcturus stream, Hercules stream, Gould's Belt, Cygnus complex, Helmi stream, Monoceros ring, etc).

Finally, the term 'bridge' is a misnomer (not connecting two long arms – see Section 4). The term 'finger' is a misnomer (not a physical entity – see Section 5). The term 'Local Arm' is a misnomer (not having the tracer offsets as seen in long spiral arm – see Section 1); it should be better known as a small 'Local Armlet'.

**Acknowledgements.** The figure production made use of the PGPLOT software at NRC Canada in Victoria. I thank an anonymous referee for useful, careful, insightful and historical suggestions.

**Table 1. Sources in the small Local Arm: quadrant, extant, pitch, shape**

| Sources | Galactic Quadrant | Max. dist. from Sun (kpc) | Pitch Angle from l=90º (º) | Note | Reference |
|---|---|---|---|---|---|
| Cepheids groups | all | 2 | 0 | random | Genovali et al (2014, Fig.10) |
| Cepheids groups | IV | 3 | 0 | random | Chené et al (2013, Fig.7) |
| HI gas | I, III | 3, 2 | -20 | - | Cohen et al (1980, Fig.3) |
| HII regions | III, IV | 3, 2 | +3 | 'branch' to Carina arm at l=282º | Xu et al (2013, Fig. 12b) |
| HII regions | I, III | 3,3 | -3 | 'branch to Perseus arm | Hou & Han (2014, Fig.5d) |
| HII regions | I | 2 | -8 | 'cylinder' | Uyaniker et al (2001, Fig.1, Tab.2) |
| HII regions | I | 2 | -20 | - | Russeil (2003, Fig.5) |
| Molecular clouds (CO) | I | 3 | -20 | 'branch' to Sagitt. arm | Jacq et al (1988, Fig.6) |
| Molecular clouds (CO) | II | 1 | 0 | Gould Belt layer + Cam OB1 layer | Du et al (2017, Fig.7) |
| O-B stars | all | 1 | 0 | random | Bobylev & Bajkova (2014, Fig.2) |
| O-B stars | I | 3 | -14 | - | Bobylev & Bajkova (2015, Fig.4) |
| Open clusters | all | 2 | 0 | random | Dib et al (2018, Fig.13) |
| Open clusters | all | 2 | 0 | random | Reddy et al (2016, Fig.7) |
| Open clusters | all | 1 | 0 | random | Piskunov et al (2006, Fig.4) |
| Open clusters | all | 3 | 0 | random | Lin & Chen (2014, Fig.1) |
| Open clusters | III | 4 | -12 | 'stellar bridge' to Perseus arm | Giorgi et al (2015, Fig.15) |

at l=225-275

| Tracer | Quadrant | Max. dist. (kpc) | z (pc) | Notes | Reference |
|---|---|---|---|---|---|
| Open clusters - young | III | 2 | -30 | - | Carraro et al (2015, Fig. 1) |
| Open clusters - young | IV | 1 | 0 | random | Carraro et al (2017, Fig. 15) |
| Open clusters – Old | all | 2 | 0 | 'ring around the Sun' | Schmeja et al (2014, Fig.6) |
| Pulsars | I,II,III,IV | 3, 1, 2, 1 | +3 | - | Yao et al (2017, Fig.9, tab.1) |
| Pulsars | all | 3 | 0 | random | Olausen & Kaspi (2014, Fig.2) |
| Trig. Masers | I, III | 4, 2 | -12 | - | Xu et al (2016, Fig. 2) |
| Trig. Masers | I | 4 | -18 | 'spur' to Sagit. arm at l=50° | Xu et al (2016, Fig.2, dots) |
| Trig. Masers | I, III | 4, 2 | -13 | - | Reid et al (2014, Fig.1) |
| Trig. Masers | I, III | 3, 2 | -10 | - | Xu et al (2013, Fig. 10, tab. 11)) |
| Trig. Masers | I, III | 4, 2 | -12 | - | Reid (2012, Fig.3, Fig.4) |
| Trig. Masers | I, III | 4, 2 | -28 | - | Reid et al (2009, Fig.2) |
| Trig. Masers | I | 4 | -20 | - | Chibueze et al (2014, Fig.5) |

Note: In a row, if more than 1 galactic quadrant is covered, then the max. distance is given accordingly for each quadrant.

Table 2. Sources in the small local arm (59º < l < 270º), with a measured trigonometric distance.

| Name | Gal. Long. (o) | Gal. Lat. (o) | Distance (kpc) See Note 1 | Syst. $V_{lsr}$ (km/s) | Reference |
|---|---|---|---|---|---|
| G059.47-00.18 | 059.5 | -0.2 | 1.87 ±0.09 | 26 | Xu et al (2016) |
| G059.78+00.06 | 059.8 | +0.1 | 2.16 ±0.07 | 25 | Reid et al (2014) |
| G059.83+00.67 | 059.8 | +0.7 | 3.95 ±0.34 | 34 | Xu et al (2016) |
| G069.54-00.97 | 069.5 | -1.0 | 2.46 ±0.07 | 12 | Reid et al (2014) |
| G071.52-00.38 | 071.5 | -0.4 | 3.61 ±0.18 | 11 | Xu et al (2016) |
| G074.03-01.71 | 074.0 | -1.7 | 1.59 ±0.05 | 5 | Reid et al (2014) |
| G074.57+00.84 | 074.6 | +0.8 | 2.72 ±0.30 | -1 | Xu et al (2016) |
| G075.76+00.33 | 075.8 | +0.3 | 3.51 ±0.29 | -9 | Reid et al (2014) |
| G075.78+00.34 | 075.8 | +0.3 | 3.83 ±0.40 | 1 | Reid et al (2014) |
| G076.38-00.61 | 076.4 | -0.6 | 1.30 ±0.09 | -2 | Reid et al (2014) |
| G078.12+03.63 | 078.1 | +3.6 | 1.64 ±0.06 | -4 | Reid et al (2014) |
| G078.88+00.70 | 078.9 | +0.7 | 3.33 ±0.24 | -6 | Reid et al (2014) |
| G079.73+00.99 | 079.7 | +1.0 | 1.36 ±0.11 | -3 | Reid et al (2014) |
| G079.87+01.17 | 079.9 | +1.2 | 1.61 ±0.07 | -5 | Reid et al (2014) |
| G080.79-01.92 | 080.8 | -1.9 | 1.61 ±0.11 | -3 | Reid et al (2014) |
| G080.86+00.38 | 080.9 | +0.4 | 1.46 ±0.09 | -3 | Reid et al (2014) |
| G081.75+00.59 | 081.8 | +0.6 | 1.50 ±0.08 | -3 | Reid et al (2014) |
| G081.87+00.78 | 081.9 | +0.8 | 1.30 ±0.07 | +7 | Reid et al (2014) |
| G090.21+02.32 | 090.2 | +2.3 | 0.67 ±0.02 | -3 | Reid et al (2014) |
| G092.67+03.07 | 092.7 | +3.1 | 1.63 ±0.05 | -5 | Reid et al (2014) |
| G105.41+09.87 | 105.4 | +9.9 | 0.89 ±0.05 | -10 | Reid et al (2014) |
| G107.29+05.63 | 107.3 | +5.6 | 0.78 ±0.06 | -11 | Reid et al (2014) |
| G108.18+05.51 | 108.2 | +5.5 | 0.78 ±0.09 | -11 | Reid et al (2014) |
| G109.87+02.11 | 109.9 | +2.1 | 0.70 ±0.04 | -7 | Reid et al (2014) |
| G121.29+00.65 | 121.3 | +0.6 | 0.93 ±0.04 | -23 | Reid et al (2014) |
| G176.51+00.20 | 176.5 | +0.2 | 0.96 ±0.03 | -17 | Reid et al (2014) |
| G203.32+02.05 | 203.3 | +2.1 | 0.74 ±0.06 | 8 | Xu et al (2016) |
| G209.00-19.38 | 209.0 | -19.4 | 0.41 ±0.01 | +3 | Reid et al (2014) |
| G213.70-12.60 | 213.7 | -12.6 | 0.86 ±0.02 | 10 | Xu et al (2016) |
| G232.62+00.99 | 232.6 | +1.0 | 1.68 ±0.10 | +21 | Reid et al (2014) |
| G239.35-05.06 | 239.3 | -5.1 | 1.17 ±0.08 | +20 | Reid et al (2014) |

Note 1: Taking all nearby sources, identified as located in the 'local arm' or spur - see Reid et al (2014). When needed, the published parallax (p, in mas) was converted to a distance (D, in kpc) through the equation D = 1/p.

**Table 3. Statistical averages of parameters, for the small Local Arm**

| Galactic Quadrant | Max. distance from Sun (kpc) | Pitch angle, from l=90° (°) | No. of tracers |
|---|---|---|---|
| I | 3.8 ± 0.2 | -16.1 ± 2.4 | 7 (only trig. masers) |
| II | 1.2 ± 0.2 | - | 7 (only trig. masers; see Figure 1) |
| III | 2.0 ± 0.3 | -15.0 ± 3.3 | 5 (only trig. masers) |
| IV | 0 | - | - (only trig. masers; see Figure 1) |
| | | | |
| I | 2.3 ± 0.3 | -5.5 ± 2.2 | 15 (all tracers, excluding trig. masers) |
| II | 1.8 ± 0.3 | +0.3 ± 0.3 | 10 (all tracers, excluding trig. masers) |
| III | 2.3 ± 0.3 | -4.2 ± 2.6 | 14 (all tracers, excluding trig. masers) |
| IV | 1.9 ± 0.3 | +0.5 ± 0.3 | 12 (all tracers, excluding trig. masers) |
| | | | |
| I,II,III,IV | 2.2 ± 0.5 | 0.0 ± - | average for all Cepheids groups |
| I,II,III,IV | 2.5 ± 0.3 | -7.0 ± 5.0 | average for all HII regions |
| I,II,III,IV | 2.5 ± 0.5 | -20.0 ± - | average for all HI gas |
| I,II,III,IV | 2.0 ± 1.0 | -10.0 ± 10.0 | average for all molecular clouds |
| I,II,III,IV | 1.4 ± 1.0 | -7.0 ± 7.0 | average for all O-B stars |
| I,II,III,IV | 2.1 ± 0.5 | -5.3 ± 4.0 | average for all open clusters |
| I,II,III,IV | 2.0 ± 0.5 | +1.5 ± 1.5 | average for all pulsars |
| | | | |
| I,II,III,IV | 3.1 ± 0.3 | -16.1 ± 2.4 | average for all trigon. masers |

**Figure captions.**

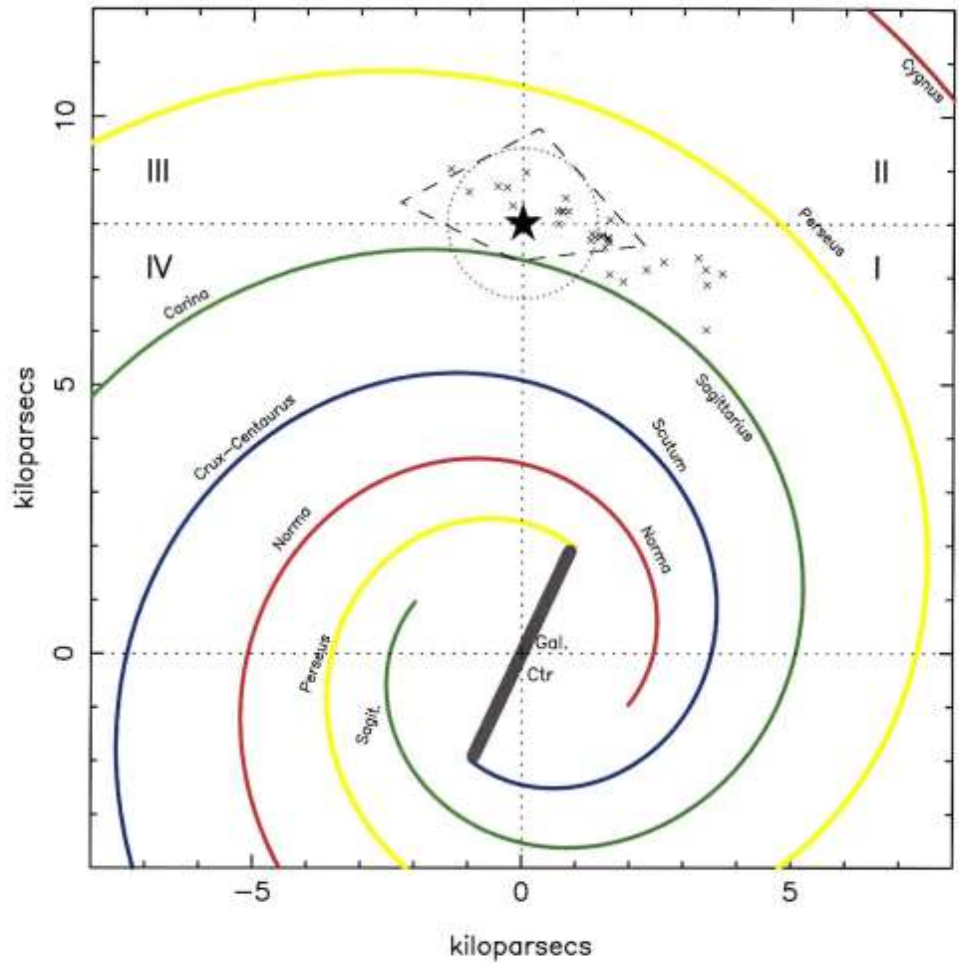

Figure 1. A view of the Milky Way's spiral arms (each arm in a different color), seen from above the galactic plane. The Sun is sketched as a star at 8.0 kpc from the Galactic Center (at 0,0). The nearest spiral arms are Perseus (yellow) and Sagittarius (green). The Galactic quadrants are shown: I (right bottom), II (right top), III (left top), and IV (left bottom). Each trigonometric maser is shown as a 'x'. The mean extent of the Local Arm, using all tracers (except the trigonometric masers), is sketched inside the 4 black dashed lines; this sketch simply joins the mean extent in 4 equidistant directions. The 4-arm model employed here is that of Vallée (2017b) and Vallée (2017d). A circle of about 1.5-kpc from the Sun is shown (blue dots).

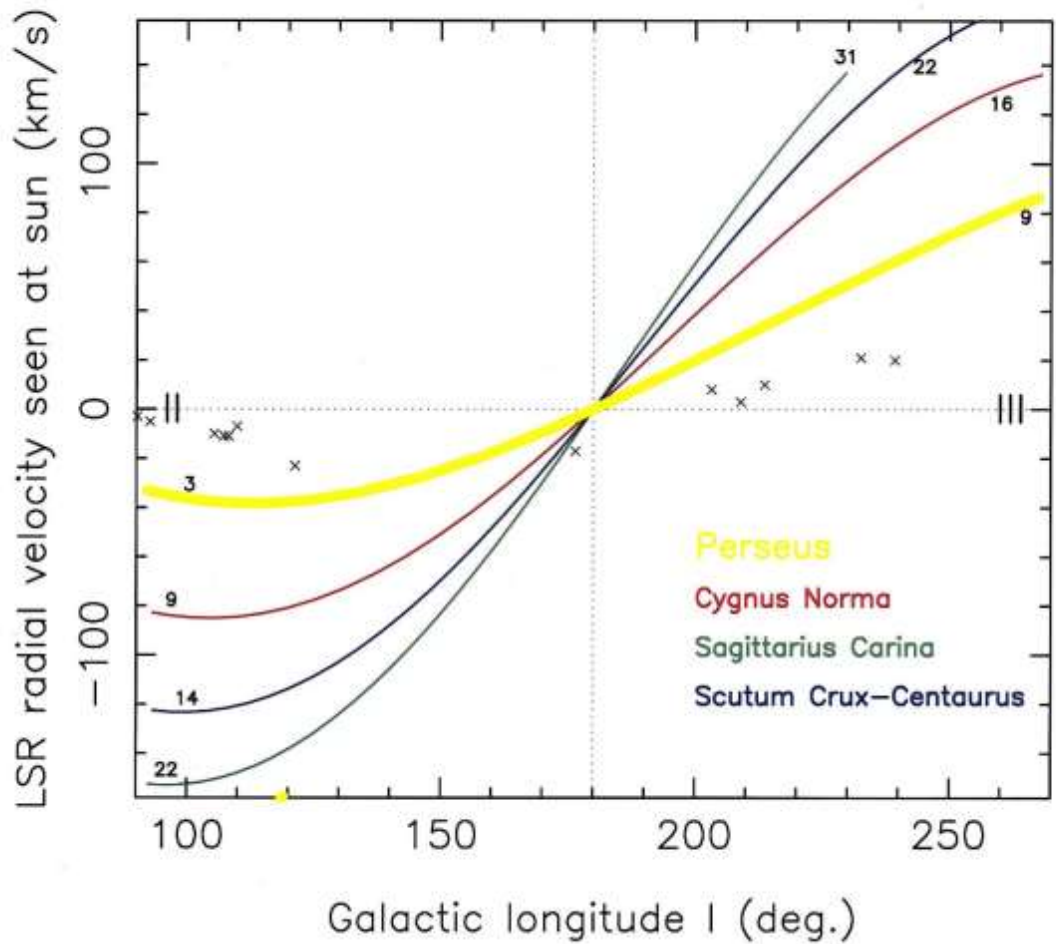

Figure 2. A view of the Milky Way's spiral arms (each arm in a different color), seen in radial velocity (vertical axis) at each galactic longitude (horizontal axis). The nearest spiral arms are Perseus (yellow) and Cygnus (red). Two Galactic quadrants are shown: II (left), and III (right). The Sun has a 0 radial velocity. Each trigonometric maser is shown as a 'x' sign; all have a radial velocity within 20 km/s of the Sun. The 4-arm model employed here is that of Vallée (2017b) and Vallée (2017d).

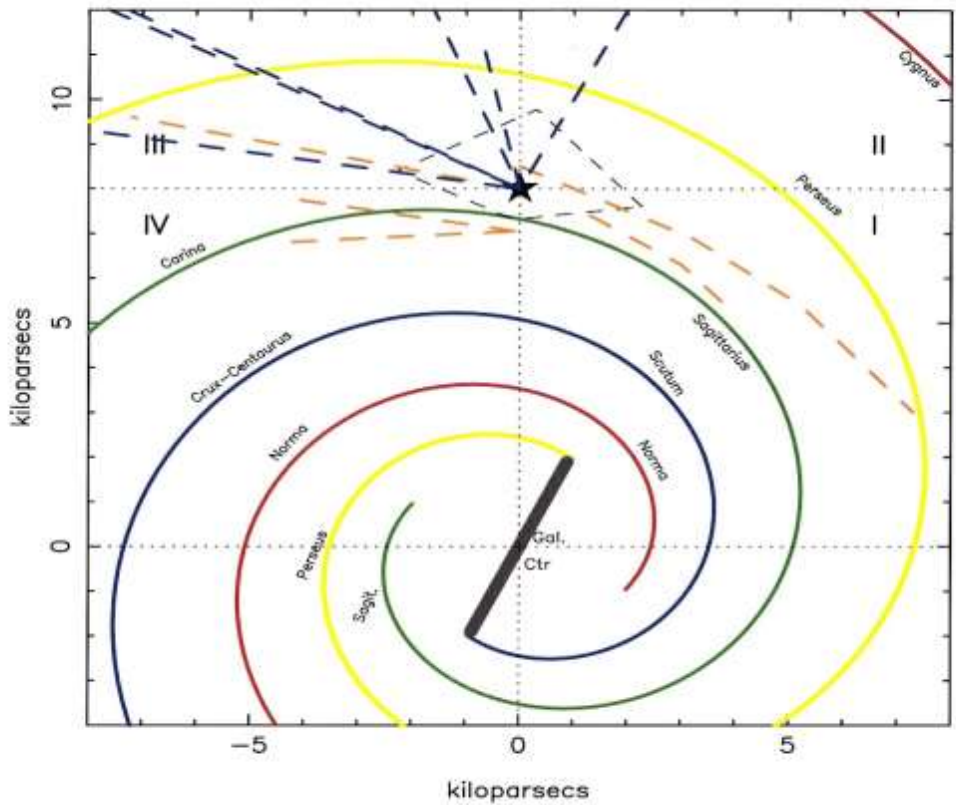

Figure 3. A view of the Milky Way's spiral arms (each arm in a different color), seen from above the galactic plane. The Sun is sketched as a star at 8.0 kpc. Same details as in Figure 1. Various 'bridges' near the Sun (orange dashed lines) have been proposed; 5 of these are shown here:  a maser 'bridge' to the Sagittarius arm (Xu et al 2016) and a model bridge to the Perseus arm (Xu et al 2013) in GQ I;  an open star cluster bridge to the Perseus arm, near l=258° (Giorgi et al 2015) in GQ III; and a forked open star cluster bridge to the Sagittarius arm (Carraro et al 2017) in GQ IV. Various 'fingers' away from the Sun  (blue dashed lines, in GQ II and III) have been proposed; 6 fingers are shown here:   a line of O stars near l=190° (Bobylev and Bajkova 2015),  a line of young open star clusters near 240°  (Carraro 2015); a line of CO clouds near l=260° (Vazquez et al 2010), a line of HI toward l=150°, another line towards l=200° and a line towards l=235° (Levine et al 2006). There are gaps in observational data as one goes away from the Sun along each 'bridge' or 'finger', so in practice these bridges or fingers may not be long, nor reach the nearest spiral arms.